# Parsec-Scale Radio Structure of the Double Active Nucleus of NGC 6240

Jack F. Gallimore[1], Department of Physics, Bucknell University, Lewisburg, PA 17837 USA, jgallimo@bucknell.edu

Robert Beswick, The University of Manchester, Jodrell Bank Observatory, Macclesfield, Cheshire SK11 9DL, United Kingdom, rbeswick@jb.man.ac.uk

## ABSTRACT

NGC 6240 is a luminous infrared galaxy that, based on recent *MERLIN* and *CHANDRA* observations, appears to host a pair of active galactic nuclei. We present new, multiepoch and multifrequency radio observations of NGC 6240 obtained using the Very Long Baseline Array (VLBA). These observations resolve out all of the diffuse radio continuum emission related to starburst activity to reveal three compact sources, two associated with the infrared and X-ray nuclei: components N1, the northern nucleus, and S, the southern nucleus; and component S1, a weak point source lying northeast of component S. The radio continuum properties of N1 and S both resemble those of compact radio sources in Seyfert nuclei, including an inverted spectrum at low frequencies and high brightness temperatures. Component N1 further resolves into a 9 pc linear source, again resembling compact jets in some well-studied Seyfert nuclei. The inverted radio spectrum is most likely caused by free-free absorption through foreground plasma with emission measures of order $10^7$ cm$^{-6}$ pc, consistent with the X-ray spectroscopy which infers the presence of a foreground, Compton thick absorber. Synchrotron self-absorption may also contribute to the flat spectrum of component S. Component S1 better resembles a luminous radio supernova such as those detected in Arp 220, although we cannot specifically rule out the possibility that S1 may be ejected material from the southern active nucleus. We demonstrate that the light curve of S1 is at least consistent with current radio supernova models, and, given the high star-formation rate in the nuclear region and the (possible) proclivity of infrared merger galaxies to produce luminous RSNe, it is not surprising that one such supernova is detected serendipitously in the VLBA observations.

---

[1] Presently on leave at the Space Telescope Science Institute, 3700 San Martin Dr., Baltimore, MD 21218, gallim@stsci.edu.

# 1 Introduction

Luminous infrared galaxies (LIRGs) are invariably strongly disturbed, presumably merging systems (Soifer et al. 1984), and it is thought that the unusually large infrared luminosity may be supported by either a resulting starburst or AGN. The central regions of these violent mergers are very dusty, hence the high infrared luminosity, and the copious dust heavily attenuates the nuclear emission from mid-IR through X-ray wavelengths (Sanders et al. 1988; Genzel et al. 1998). What makes the question more intriguing is the observation that the co-moving volume density of LIRGs compares to that of QSOs in equal luminosity bins. This result led Sanders et al. (1988) to argue that violent mergers may be the source for all QSOs, or, equivalently, that all QSOs are the final phase of a violent merger. The QSOs are enshrouded by dust when they first form and appear to us as a ULIRG. As the AGN destroys or blows away its dusty shroud the source takes on the characteristics of a classical optical QSO. Later efforts argued however that the final state of most of these mergers will probably be L* elliptical galaxies hosting lower-luminosity, Seyfert-like AGNs (Colina et al. 2001b; Tacconi et al. 2002).

NGC 6240 ($D = 97$ Mpc for $H_0 = 75$ km s$^{-1}$ Mpc$^{-1}$; $1'' = 470$ pc; de Vaucouleurs et al. 1991) is a well-studied merging pair of galaxies (Fosbury & Wall 1979; Fried & Schulz 1983). As is common for violently merging systems, NGC 6240 produces a large far infrared luminosity, $L_{IR} \sim 6 \times 10^{11}$ $L_\odot$ (Soifer et al. 1984; Wright, Joseph, & Meikle 1984; Sanders et al. 1988). This luminosity is roughly half that of the prototype ULIRG Arp 220 (e.g., Sanders et al. 1988), another violent merger, and compares to the bolometric output of QSOs. Although NGC 6240 appears to be in a relatively early stage of merging (Tacconi et al. 1999; Tecza et al. 2000; Genzel et al. 2001), there is strong evidence, particularly from X-ray (Komossa et al. 2003) and radio (Beswick et al. 2001) studies, that the merger may have already triggered activity in both of the parent nuclei.

With the angular resolution afforded by the Very Large Array (VLA), the nuclear radio structure of NGC 6240 comprises three compact sources, two of which (components N1 and S, following the labeling scheme of Beswick et al. 2001) mark the positions of the parent nuclei, and extended amorphous structure that appears to be associated with the starburst or a starburst-driven superwind (Colbert, Wilson, & Bland-Hawthorn 1994; Beswick et al. 2001). At VLA and MERLIN resolutions, all of the compact radio sources have steep centimeter-wave spectra. As shown in
Figure 1, MERLIN observations resolve out most of the diffuse, extended radio continuum, but the radio nuclei N1 and S remain relatively bright and compact (Tacconi et al. 1999; Beswick et al. 2001). Although the southern nucleus displays some faint, lobe-like extensions, both nuclei remain unresolved by the 50 milliarcsecond (mas) MERLIN beam, limiting the sizes to < 25 pc and brightness temperatures > $10^5 - 10^6$ K.

The direct, integrated X-ray spectrum of NGC 6240 is dominated by soft, thermal emission characteristic of a starburst (Turner et al. 1997), but there is also a hard X-ray component compatible with a highly absorbed AGN (Vignati et al. 1999). Iwasawa &

Comastri (1998) reported the detection of a Fe K$\alpha$ line based on sensitive ASCA spectroscopy (see also Tanaka, Inoue, & Holt 1994; Kii et al. 1997). The high equivalent width of the line is strong evidence for a hidden AGN (Krolik & Kallman 1987; Awaki et al. 1991). Subsequent Chandra observations detected and resolved both nuclei as compact, hard (5 – 8 keV) X-ray sources (Komossa et al. 2003). Taken with the MERLIN observations, the conclusion is that both of the parent nuclei are active, and Komossa et al. dubbed NGC 6240 a "binary AGN."

We followed up our MERLIN studies with Very Long Baseline Array[2] (VLBA) observations of the two nuclei of NGC 6240. Our primary goal was to confirm the compact, AGN-like nature of the radio nuclei and to look for evidence of a compact radio jet such as is seen in Seyfert galaxies or CSS / GPS sources. These observations led to the serendipitous discovery of a faint, compact radio source, which we call "component S1," between the parent nuclei that may be a fading radio supernova (RSN) (Gallimore & Beswick 2000). We report in this study the results of these VLBA observations and an additional follow-up MERLIN observation. The following sections present, respectively, the details of the multi-epoch VLBA and MERLIN observations and data reduction, the imaging results for the two radio nuclei, the imaging results for the candidate radio supernova, and a summary of the key results.

## 2 Observations and Data Reduction

### 2.1 VLBA Observations

We first observed NGC 6240 with all 10 antennas of the VLBA on August 16, 1999. Observations were made through the S/X dichroic, giving us simultaneous continuum observations at 2.3 and 8.4 GHz and a bandwidth of 32 MHz at each frequency. Particularly at 8.4 GHz, the ~ 1.5″ nuclear region spans many VLBA beams, and so we alternated pointed scans of the northern (component N1) and southern (component S) nuclei in an attempt to alleviate potential bandwidth-smearing and time-averaging smearing. The total times on-source were 135 minutes for each nucleus. The source observations were interleaved every three minutes with a two minute scan of the phase reference source J1641+0129, located ~ 1° from NGC 6240.

After reduction, we discovered 2.4 GHz radio emission from component S1 which we interpreted as a candidate radio supernova (Gallimore & Beswick 2000). We therefore followed up these pilot observations with additional 2.4 and 1.7 GHz observations with the main goal of following the radio light curve of the supernova candidate. The salient parameters of the observations are summarized in Table 1. No sources brighter than 0.36 mJy (3$\sigma$) were detected on the 8.4 GHz images.

---

[2] The VLBA is operated by the National Radio Astronomy Observatory. The National Radio Astronomy Observatory is a facility of the National Science Foundation operated under cooperative agreement by Associated Universities, Inc.

Data reduction entailed standard procedures using the software package AIPS provided by the NRAO. After initial editing of obviously faulty visibilities, the data were corrected for ionospheric delays based on ionosphere electron content maps provided by the Jet Propulsion Laboratory. The flux scale was set by an *a priori* amplitude calibration based on system temperatures and gain corrections provided as tables attached to the VLBA data. Phases were aligned across IF channels using the internal pulse calibration system, and the single-band delay was referenced to the bright radio source 3C 345. Finally, the rates, delays, and phases of the NGC 6240 data were calibrated based on observations of the nearby phase reference source J1651+0129. The phase reference source was further self-calibrated both for phase and amplitude corrections, and the corrections were transferred to the NGC 6240 visibilities.

The NGC 6240 data were then averaged in time and frequency. The first epoch, 2.4 GHz datasets that were observed at two different positions were shifted to a common position and combined. Images were produced by standard Fourier inversion and CLEAN deconvolution tasks in AIPS. Visibilities were naturally weighted during the Fourier transform to ensure the best sensitivity. Table 2 summarizes the basic properties of the resulting images.

The resulting images are shown as contour maps in Figures 2 – 7. Owing to the different image sensitivities, it was not practical to use identical contour levels; otherwise, the resulting maps would include either excessive noise contours (on lower SNR images) or loss of detail (on higher SNR images). The strategy we chose included contours at the $\pm 3\sigma$ level scaled by factors of $1.5^n$, where $n$ is the contour number. The resulting contour levels are provided in Table 3.

## *2.2 MERLIN Observations*

We made two separate 5 GHz continuum observations of NGC 6240 with the MERLIN[3] telescope based at the Jodrell Bank Observatory. The first observation was an 8-hour track taken on 7 Dec 1997, and the details of this observation and a contour map were previously published by Tacconi et al. (1999). The second observation was taken on 18 May 2000 (following the discovery of component S1). Data reduction entailed primary editing and phase and amplitude calibration using the "dprocs" software local to Jodrell Bank. Subsequent editing and self-calibration were performed in AIPS. The resulting image, displayed in
Figure 1, was generated using standard Fourier inversion and CLEAN-ing techniques within AIPS. The RMS sensitivity of the second MERLIN image is 0.17 mJy beam$^{-1}$.

---

[3] MERLIN is a UK National Facility operated by the University of Manchester at Jodrell Bank Observatory on behalf of the Particle Physics & Astronomy Research Council (PPARC)

# 3  The Northern Nucleus (Component N1)

Radio component N1 shows a clear east-west elongation at both 1.7 and 2.4 GHz bands and at every epoch. We analyzed the structure by taking intensity-weighted moments of the images using a modified version of the AIPS task "MOMFT." The results are provided in Table 4. For the purpose of brevity, all coordinates are listed as seconds for right ascension and arcseconds for declination, offset from J2000 coordinates $16^h\ 52^m$, $+2°\ 24'$. Note that the astrometric errors listed in Table 4 are formal statistical errors based on the signal to noise ratio of the detections and do not include a contribution from systematic uncertainties resulting from errors in phase connection or the VLBA correlator model. Given the proximity of the calibrator to the target, we expect that the systematic uncertainty of the astrometry is around 1 mas (see, for example, Beasley & Conway 1995; Reid et al. 1999; Dhawan, Mirabel, & Rodríguez 2000) . The measured standard deviations, 0.6 mas in right ascension and 1.5 mas in declination, are compatible with the predicted uncertainty.

The VLBA, VLA, and MERLIN continuum measurements are compared in Figure 17. Most of the diffuse, steep-spectrum emission detected by the VLA (Carral, Turner, & Ho 1990; Colbert et al. 1994) resolves out in the MERLIN and VLBA images. The average VLBA spectrum is inverted with index $\alpha \cong +1$ ($S_\nu \propto \nu^\alpha$). Some extended emission may resolve out at 2.4 GHz and the actual spectrum may be more steeply inverted. The spectrum must turn over between 1.6 and 8.4 GHz given the non-detection at 8.4 GHz.

The extent of N1 is about 9 pc (20 mas). The contour images hint at an apparent change of structure, particularly between the second and third epochs, of which the images have comparable sensitivity. The epoch 2 images at both frequency bands appear to be more compact and the epoch 3 images appear more extended, with perhaps separate "lobes" appearing on the 2.4 GHz image. We viewed the structural details skeptically owing to the faintness of the extended emission. As a check, we convolved the images to a common beam size (only slightly larger than the nominal beam sizes) and, for each frequency band, subtracted the epoch 2 image from the epoch 3 image. The resulting difference maps showed no significant structure above the level of the map noise. The structural differences appearing on the contour maps are most likely an artifact of the low signal-to-noise of the lower contour levels.

N1 broadly resembles radio sources found in high luminosity Seyfert nuclei. Averaged over the angular extent of the source, the brightness temperatures exceed $7 \times 10^6$ K, arguing for an AGN origin; in contrast, the average brightness of star-forming regions is $\sim 10^4$ K (Condon 1992). The extended structure resembles compact jets found in Seyfert galaxies or GPS sources. The radio power of N1, $6 \times 10^{21}$ W Hz$^{-1}$ at 2.4 GHz, falls in the range appropriate for Seyfert nuclei (Ulvestad & Wilson 1989; Morganti et al. 1999), but four orders of magnitude less luminous than CSS and GPS sources (O'Dea 1998).

The inverted radio spectrum may result from a combination of synchrotron self-absorption and free-free absorption. Synchrotron self-absorption only affects compact radio sources with high brightness temperature (Kellermann & Pauliny-Toth 1969). The signal-to-noise ratios of the VLBA sources are too low to constrain the peak brightness temperatures sufficiently to evaluate the impact of synchrotron self-absorption on any single compact source, but, based on image moment analysis, compact (unresolved) structures make up less than half of the VLBA radio continuum emission from the northern nucleus. Therefore, given the low average brightness temperature of the source as a whole, it seems that free-free absorption through foreground, ionized gas has a greater impact on the spectrum. Since there is no evidence for free-free absorption in the VLA/MERLIN spectrum, the size of the absorbing medium must be quite compact, between 9 – 25 parsecs in extent. A small circumnuclear disk or torus, such as is posited for AGN unifying schemes, might produce the requisite free-free absorption (Antonucci 1993; Pedlar et al. 1998; Gallimore et al. 1999; Ulvestad et al. 1999). For comparison, the X-ray spectrum and luminosity also match the characteristics of a heavily obscured Seyfert nucleus (Kii et al. 1997; Iwasawa & Comastri 1998; Vignati et al. 1999; Matt et al. 2000; Lira et al. 2002; Komossa et al. 2003), although the X-ray luminosity may be as great as $10^{45}$ ergs s$^{-1}$, comparable to QSOs (Iwasawa & Comastri 1998; Lira et al. 2002). Assuming an intrinsic spectral index $\alpha = -0.7$ and purely foreground absorption, $\tau_{ff}(\nu = 1.6$ GHz$) \sim 1.1$. The free-free opacity at GHz frequencies is directly proportional to the emission measure, EM; the proportionality is approximated by:

$$\tau_{ff} = 3.278 \left(\frac{T_e}{10^4 \text{ K}}\right)^{-1.35} \left(\frac{\nu}{1 \text{ GHz}}\right)^{-2.1} \left(\frac{\text{EM}}{10^7 \text{ cm}^{-6} \text{ pc}}\right) \quad (1)$$

(Altenhoff et al. 1960; Mezger & Henderson 1967). Further assuming that the free-free absorption arises from a purely hydrogen plasma at $T_e = 10^4$ K, EM $\sim 1 \times 10^7$ cm$^{-6}$ pc.

It is straightforward to interpret N1 as a radio jet originating from a heavily obscured Seyfert nucleus or other sort of intermediate luminosity AGN (Iwasawa & Comastri 1998; Lira et al. 2002), but can we rule out a significant contribution from RSNe? Discussed below in Section 5.1, at most we would expect to detect 1 or 2 luminous RSNe at any given epoch throughout the whole nuclear region of NGC 6240. Given that most of the molecular gas mass and star formation within NGC 6240 falls between the two nuclei (Tacconi et al. 1999), it seems unlikely that the bulk of the star formation would take place within a few pc of one of the compact VLBA sources. It seems more likely that the VLBA-scale radio emission associated with N1 is produced by AGN-powered outflow.

## 4  The Southern Nucleus (Component S)

In contrast to component N1, radio component S is unresolved by the VLBA; the implied source size is < 2 pc. The radio continuum properties are otherwise similar to those of N1; see Figure 18. Like the northern nucleus N1, S has an inverted spectrum between 1.7 and 2.4 GHz (spectral index $\alpha = 3.6$). The radio power at 2.4 GHz is $3.9 \times 10^{21}$ W Hz$^{-1}$, and the brightness temperature exceeds $18 \times 10^6$ K.

Based on the X-ray evidence and the radio luminosity, it appears that radio component S also associates with a hidden, intermediate luminosity AGN, and the radio and X-ray properties fall within the range of parameters filled by Seyfert nuclei. The parsec-scale radio spectrum is steeply inverted, but the MERLIN spectrum is steep, and so the VLBA radio source appears to be absorbed on scales < 25 pc. Given the compactness of the source, synchrotron self-absorption may affect the spectrum here, but we cannot specifically rule out free-free absorption. Making the same assumptions used above for the analysis of component N1, a free-free opacity of ~ 3.6, corresponding to an emission measure of ~ $3 \times 10^7$ cm$^{-6}$ pc, would be required to explain the steeply inverted spectrum.

It is interesting to note that there is $H_2O$ megamaser emission associated specifically with component S (Hagiwara, Diamond, & Miyoshi 2002, 2003), indicating the presence of dense molecular gas within ~ 3 pc of the nucleus. The central X-ray source would ionize the inner surface of the molecular maser clouds and produce free-free absorption. Based on models describing the properties of X-ray irradiated molecular gas (hereafter, "maser models"), the expected free-free opacity is given by

$$\tau_{ff}(\nu) \approx 0.4 \left[ \left( \frac{L_X}{10^{43} \text{ ergs s}^{-1}} \right) \left( \frac{\text{pc}}{R} \right)^2 \right]^{1.1} \left( \frac{n_e T_e}{10^{11} \text{ K cm}^{-3}} \right)^{-0.1} \left( \frac{\nu}{10 \text{ GHz}} \right)^{-2.1} \qquad (2)$$

where $L_X$ is the 1—100 keV X-ray luminosity, and $R$ is the distance between the X-ray source and the molecular gas (Neufeld, Maloney, & Conger 1994; Neufeld & Maloney 1995). We can use the upper limit on the free-free opacity ($\tau_{ff} \leq 3.6$ at 1.7 GHz, derived above) and the distribution of $H_2O$ masers ($R = 3$ pc) to constrain the X-ray luminosity: $L_X \leq 2.5 \times 10^{43}$ ergs s$^{-1}$ assuming an electron pressure $n_e T_e = 10^{11}$ K cm$^{-3}$, which is appropriate for maser clouds (Neufeld et al. 1994). Note that the formal dependence on the electron pressure is weak.

The observed X-ray luminosity of either nucleus is of order $10^{41}$ ergs s$^{-1}$ (from Komossa et al. 2003 and adjusted to a distance of 97 Mpc), two orders of magnitude below the luminosity of component S based on the maser models. This result supports the view that we see only a scattered component of the X-ray source and not the source directly at energies below 10 keV. Models for the soft and hard X-ray spectra of NGC 6240 moreover require the presence of a Compton-thick absorber and hard X-ray sources with total luminosity $L_X \sim 10^{43}$—$10^{44}$ ergs s$^{-1}$ (Vignati et al. 1999; Ikebe et al. 2000; Matt et al. 2000). On the other hand, if instead the spectrum is more severely affected by synchrotron self-absorption, then the predicted X-ray luminosity would be correspondingly reduced. Turning the argument around, to maintain self-consistency with the maser model and ensure $L_X \geq 10^{43}$ ergs s$^{-1}$, the free-free opacity at 1.7 GHz must exceed ~ 2.

# 5 The Nature of Component S1

The nature of component S1 is not as clear as N1 or S given its offset position from either nucleus. We initially interpreted this source as a candidate RSN since it was not clearly detected in earlier MERLIN data (Gallimore & Beswick 2000). On the other hand, its proximity to component S and a possible extension of that source raises the possibility that S1 may simply be radio ejecta associated with the southern nucleus. We consider each possibility, as well as the remaining possibility that S1 may simply be a background source, in turn.

## 5.1 Component S1 as a Background Radio Source

The question in this case is what is the probability that a background radio source brighter than 0.5 mJy would, by chance, be located between the nuclei of NGC 6240? Based on VLA observations, the relevant number density of radio sources on the sky is $n(S_{1.4} \geq 0.5 \text{ mJy}) \sim 1.5 \times 10^{-5}$ (arcsec)$^{-2}$ (Windhorst et al. 1985). For comparison, only one source brighter than 0.5 mJy was detected in deep 1.4 GHz MERLIN observations of a $10'' \times 10''$ region centered on the Hubble Deep Field (Muxlow et al. 1999), and that source resolves below 0.4 mJy peak in EVN[4] observations (Garrett et al. 2001). The MERLIN 5 GHz nuclei of NGC 6240 are separated by ~ 1.5", and so the probability of detecting a background radio source with the VLA is of order 1 in 30000, and the probability of detecting a background source with MERLIN is of order 1 in 160000 (accepting small number statistics for that particular calculation). The probability of a VLBA detection of a background source becomes even more remote, since MERLIN sources will commonly resolve out on VLBI-scale baselines. Of course, we cannot specifically rule out this interpretation for component S1, but the probability that S1 is a background radio source is miniscule.

## 5.2 Component S1 as Radio Ejecta associated with the Southern Nucleus

Compact radio sources associated with radio-emitting ejecta are observed in a large number of Seyfert nuclei and other low – intermediate luminosity AGNs (Ulvestad & Wilson 1989; Kukula et al. 1995; Morganti et al. 1999; Thean et al. 2001). Interpreting component S1 as ejecta from component S would be consistent with the identification of component S as an AGN. This scenario might be supported by the radio structure of nucleus S in the MERLIN 5 GHz map (
Figure 1): there appears northeast of the southern nucleus an extended component, partially resolved away from the MERLIN core. However, it should be noted that the point response (beam) of these MERLIN observations extends to the northeast and also exhibits strong sidelobe structure in that direction.

---

[4] European VLBI Network, a joint facility of European, Chinese, South African and other radio astronomy institutes funded by their national research councils.

The low surface brightness features on past MERLIN images have been furthermore inconsistent with the data presented here. Looking at Beswick et al. (2001) for comparison, their MERLIN images of component S show extensions to the southeast (at 5 GHz) and to the northwest (at 1.4 GHz). Given these inconsistencies; the low surface brightness features surrounding component S should be viewed skeptically. The origin of the apparent, extended radio features ("extensions," for short) is probably an artifact of the aperture synthesis technique and owes to a combination of factors: (1) the MERLIN telescope is a sparse array with large gaps in ($u, v$) coverage, (2) NGC 6240 is located near the celestial equator (declination ~ +2°), degrading the effect of Earth-rotation synthesis on the ($u, v$) coverage, and (3), detected by the VLA, which is more sensitive to extended emission, there is a diffuse skirt of radio emission, probably associated with star formation, surrounding component S (Colbert et al. 1994; Beswick et al. 2001). Aperture synthesis imaging techniques can introduce faint, artificial clumps and extensions into such extended diffuse emission, especially when it is observed by a sparse, long baseline array such as MERLIN (Swain 1996; Cornwell, Braun, & Briggs 1999).

Further complicating this interpretation, component S1 has a fairly flat spectrum at low frequencies (consistent with $\alpha = 0$ between 1.7 – 2.4 GHz; see Table 6), but the radio ejecta of low and intermediate luminosity AGNs generally have steep spectra consistent with optically thin synchrotron radiation (cf. Morganti et al. 1999; Thean et al. 2001 and references therein). Of course, free-free absorption, perhaps through an HII region in the surrounding star forming regions, may have flattened the 1.7 – 2.4 GHz VLBA spectrum. At this point, however, the model becomes more complicated than the data allow. While we cannot specifically rule out the possibility that S1 is associated with ejecta from the southern nucleus, there is no clear connection between components S and S1 other than proximity and a dubious radio extension of component S that was detected in one of three independent MERLIN observations.

### *5.3 Component S1 as a Luminous RSN*

Component S1 has appeared on all three epochs of VLBA observations but not on earlier and later MERLIN images. The 2.4 GHz light curve of component SN has steadily decreased in the three epochs of VLBA observations, and the 1.7 GHz flux density similarly decreased between epochs 2 and 3. These data provide insufficient evidence to identify this variable radio source conclusively as a RSN, but the question of plausibility remains. Should we see luminous RSNe in NGC 6240, and are the light curves of component S1 compatible with well-studied, luminous RSNe?

To address the first question, it is convenient to scale relative to the properties of the Arp 220 RSNe. We will assume, for the sake of argument, that the nuclear region of NGC 6240 has the same proclivity for producing luminous RSNe as Arp 220. Projecting Arp 220 to the distance of NGC 6240 and accounting for the sensitivity of our VLBA images, we would have detected 7 of the 16 RSNe found by Smith et al. (1998) at the 5$\sigma$ level. Since the infrared luminosity of NGC 6240 is about half that of Arp 220 (Soifer et al. 1984; Wright et al. 1984; Sanders et al. 1988; Dopita et al. 2002), and further accounting that star-formation powers only roughly half of the infrared luminosity of NGC 6240 (Genzel et al. 1998), we should expect to detect one or two luminous RSNe at any given time in NGC 6240. Therefore, assuming that Arp 220 is representative of starbursting mergers like NGC 6240, it seems likely that we should have detected a luminous RSN during the three years of VLBA observations.

Turning to the second question, the GHz-frequency light curves of RSNe comprise an initial, rapid brightening of a few hundred days duration, peaking first at higher frequencies, followed by a power-law decline with a time scale of years (Weiler et al. 1986; Weiler, Panagia, & Sramek 1990; Montes, Weiler, & Panagia 1997; Weiler et al. 2002). The light curves have been successfully modeled by a combination of steep-spectrum synchrotron emission from the expanding supernova bubble attenuated by decreasing free-free opacity of surrounding, circumstellar material (Chevalier 1982a, 1982b; Weiler et al. 1986; Chevalier & Fransson 2001; Weiler et al. 2002). As the free-free opacity drops below unity, the light curve peaks, and then the supernova fades with the synchrotron emissivity of the supernova ejecta.

Following the analysis of the past luminous RSNe (Weiler et al. 1990; Weiler et al. 2002), the simplest model for the light curve can be expressed as:

$$S_\nu(t) = S_0 \left(\frac{\nu}{5\,\text{GHz}}\right)^\alpha \left(\frac{t-t_0}{1\,\text{day}}\right)^\beta e^{-\tau_{ff}} \quad (3)$$

where $S_0$ is the intrinsic (unobscured) 5 GHz flux density one day after detonation, $\alpha$ is the spectral index, $\beta$ is the power law index describing the decay of the synchrotron emissivity, $t_0$ is the date of the explosion, and $\tau_{ff}$ is the time-dependent free-free opacity of the circumstellar medium (CSM). The free-free opacity is given by

$$\tau_{ff}(t) = \tau_0 \left(\frac{\nu}{5\,\text{GHz}}\right)^{-2.1} \left(\frac{t-t_0}{1\,\text{day}}\right)^\delta \quad (4)$$

where $\tau_0$ is the free-free opacity of the CSM one day after detonation, and, assuming that the CSM is pre-detonation stellar wind material (with density profile $\rho \propto r^{-2}$), $\delta = \alpha - \beta - 3$ (Chevalier 1982b).

We fit this radio light curve model to our data using a non-linear chi-squared minimization algorithm; the results are provided in Table 7 and Figure 19. The formal 1σ errors listed in Table 7 were derived from the parameter covariance matrix. This model has five free parameters ($t_0$, $S_0$, $\tau_0$, $\alpha$, $\beta$), and our observations comprise five detections and three non-detections (upper limits). The non-detections at 5 GHz primarily constrain the time of maximum light at 5 GHz, and, because it coincides with detection at 2.4 GHz, the non-detection at 8.4 GHz limits the luminosity of the RSN. The constraints afforded by the upper limits are soft, and, as should be expected, there exists no unique best-fit model when all five model parameters are allowed to vary. Since our goal however is to see if the data can be described by a plausible supernova light curve, we fixed the spectrum and light curve indices to those measured for the luminous RSN 1986J, $\alpha = -0.7$ and $\beta = -1.3$ (Weiler et al. 1990). These values appear to be typical for well-studied RSNe (Weiler et al. 2002).

The model predicts a peak 5 GHz flux density of about 1 mJy, which, at the distance of NGC 6240, corresponds to a peak radio luminosity $P_5 = 10^{21}$ W Hz$^{-1}$, similar to the brightest VLBI RSNe detected in the nuclei of Arp 220 (Smith et al. 1998) and a luminous RSN detected in NGC 7469 (Colina et al. 2001a). For comparison, the range of luminosities of detected and well-studied type Ib and Ic RSNe is $10^{19.7} - 10^{21.8}$ W Hz$^{-1}$, and the range for detected type II RSNe is $10^{19.1} - 10^{22.1}$ W Hz$^{-1}$ (Weiler et al. 2002). The best-fit CSM opacity, $\tau_0 \sim 10^5$, is also comparable to the values found for type II RSNe, although the CSM opacities of type Ib and Ic RSNe tend to be somewhat lower. We conclude that the multi-frequency light curves of component S1 are at least comparable to those of well-studied, luminous RSNe. At the time of this writing, component S1 may remain bright enough at lower frequencies to follow the fading light curve.

## 6  Discussion

VLBA imaging has revealed three compact radio sources within the central 2" of NGC 6240. Two of the radio sources, components N1 and S, agree well with the positions measured on MERLIN images (Beswick et al. 2001) and appear also to coincide with compact hard X-ray sources[5] (Komossa et al. 2003). Both VLBA sources have inverted radio spectra below 2.4 GHz, suggesting strong free-free absorption on scales < 25 pc. Models of the X-ray spectra also argue for a Compton-thick obscuring medium (Vignati et al. 1999; Ikebe et al. 2000; Matt et al. 2000). Taken together, the radio and X-ray data argue that both components mark hidden, low luminosity AGNs with radio and X-ray properties comparable to those of Seyfert nuclei.

What triggered the AGNs? It is commonly thought that major mergers can trigger nuclear activity, a view that is bolstered by the disturbed appearance of QSO host galaxies (Hutchings & Neff 1988, 1992; Bahcall, Kirhakos, & Schneider 1995; Bahcall et al. 1997; Hutchings & Neff 1997; Kirhakos et al. 1999; Hutchings et al. 2002). It is difficult to simulate merger-induced fueling of AGNs and compact starbursts owing to the requisite spatial dynamic range, but the recent models of Bekki and Noguchi (Bekki & Noguchi 1994; Bekki 1995, 2000) indicate that the greatest mass deposition to the inner kpc does not occur until late in the merging process, that is, when the two parent nuclei coalesce. On the contrary, NGC 6240 is probably in an early phase of interaction (Tacconi et al. 1999; Tecza et al. 2000), and certainly the two nuclei are still discernable. It seems more likely that the parent galaxies, presumably bright, gas-rich spirals (Fosbury & Wall 1979), were already predisposed towards nuclear activity, possibly owing to disk instabilities (e.g., Shlosman, Frank, & Begelman 1989; Shlosman, Begelman, & Frank 1990). Any such instabilities may have been helped along by milder tidal effects early in the merging process (Noguchi 1988; Hernquist 1989; Mihos & Hernquist 1994; Hernquist & Mihos 1995). Alternatively, successive orbits of the parent galaxies may have triggered starburst events (Barnes & Hernquist 1996; Mihos & Hernquist 1996) which in turn may have activated, or were co-activated with, the active nuclei (Norman & Scoville 1988; Cid Fernandes & Terlevich 1995; Heckman et al. 1995; Colina et al. 1997; Williams, Baker, & Perry 1999).

---

[5] Komossa et al. do not provide astrometry for the X-ray nuclei, but they report a separation of 1.4″ between the nuclei with imaging resolution 0.9″. The separation of the VLBA radio sources is 1.5″.

Turning the argument around, there have been many paths proposed ultimately to drive nuclear activity by way of interactions, but in heavily obscured LIRGs it is difficult to identify the presence of a QSO (e.g., Genzel et al. 1998), much less lower luminosity nuclear activity. The presence of a binary AGN in NGC 6240 prompts the question of low luminosity AGN activity in other LIRGs that may either be buried in dust or swamped by a powerful starburst. It will be interesting to see whether future X-ray and VLBI studies turn up additional LIRG Seyfert nuclei and whether their presence is related to the state of the merger.

The nature of radio component S1 is more difficult to assess. It has been detected only on VLBA images taken between 1999 and 2001 and not on MERLIN images that bracket these epochs. Considering the expected supernova rate for this luminous infrared galaxy, we have interpreted this fading radio source as a RSN that is comparable in luminosity to those detected in the prototype LIRG Arp 220 (Smith et al. 1998) and a single RSN detected in the LIRG / Seyfert NGC 7469 (Colina et al. 2001a). The light curve has been sampled only sparsely, however, but we have demonstrated that the data are at least compatible with observations of other luminous, extragalactic RSNe and a reasonable light curve model.

RSNe also tend to be luminous X-ray sources, and so should component S1 have been detected by X-ray observatories? Characteristic X-ray luminosities, as measured by the Einstein, ROSAT, ASCA, and Chandra observatories in their respective energy bands, fall in the range $L_X = 10^{37} - 10^{39}$ ergs s$^{-1}$ (Schlegel 1994; Schlegel 1995; Immler & Pietsch 1998; Immler, Pietsch, & Aschenbach 1998; Immler & Pietsch 2001; Ray, Petre, & Schlegel 2001; Schlegel & Ryder 2002). More luminous examples include SN 1986J, with X-ray luminosity $\sim 2 \times 10^{40}$ ergs s$^{-1}$ (Houck et al. 1998), and the radio-luminous SN 1988Z was marginally detected by ROSAT indicating a large X-ray luminosity of $\sim 8 \times 10^{40}$ ergs s$^{-1}$ (Fabian & Terlevich 1996). Even if component S1 were as X-ray luminous as SN1988Z, it would still be very difficult to detect even by the Chandra observatory owing to its proximity to the southern nucleus. To be specific, the directly observed X-ray luminosity of component S is $9 \times 10^{41}$ ergs s$^{-1}$ (Komossa et al. 2003), an order of magnitude more luminous than SN 1988Z. Component S1 is located only 0.3″ from S, but the angular resolution of the Chandra High Resolution Camera is $\sim 0.5″$ (Kenter et al. 2000). Any supernova-like X-ray emission from component S1 would be difficult to distinguish from component S.

We have examined archival HST images which happened to be taken within two years following time of the modeled 5 GHz radio peak. There appears to be no optical counterpart to component S1, to a limiting magnitude of 22.6 at 830 nm (F814W WFPC-2 filter; March 1999) using the Vega system, or infrared counterpart, to a limiting magnitude of 18.6 at 2.22μm (F222M NICMOS filter; March 1998); note that those observations are limited largely by confusion with diffuse emission throughout the nuclear environment, and there exist no earlier, matching observations to perform subtraction. Component S1 is however located near a dense concentration of molecular gas between the two nuclei (Bryant & Scoville 1999; Tacconi et al. 1999), and the region also appears to be heavily reddened even at near-infrared wavelengths (Scoville et al. 2000; Tecza et al. 2000). Moreover, the optical maximum of a supernova generally precedes the radio maximum (Weiler et al. 2002). In the context of the supernova interpretation, it is therefore not surprising that an optical or infrared counterpart failed to appear on archival HST images.

At the time of this writing component S1 may be detectable at low frequencies, but resolving it from nuclear component S will require sensitive VLBI observations. It may be difficult to confirm the nature of component S1 as it fades, but, given an expected supernova rate of order 1 yr$^{-1}$ and a possible proclivity of mergers to produce radio luminous RSNe (Smith et al. 1998), a better test would be to search for new radio supernovae. The appearance and fading of new compact radio sources in the neighborhood of component S1 would provide strong support for the supernova model, akin to the interpretation of the compact radio sources in the nucleus of Arp 220 (Smith et al. 1998).

**Acknowledgements**

We would like to acknowledge discussions with Reinhard Genzel and Linda Tacconi which motivated this project, and we also wish to express our gratitude to Ed Colbert, the referee, whose suggestions greatly improved the clarity of the text. JFG was supported by an NRAO Jansky Fellowship during the beginning phases of this work. This research has made use of NASA's Astrophysics Data System Bibliographic Services.

**Figures**

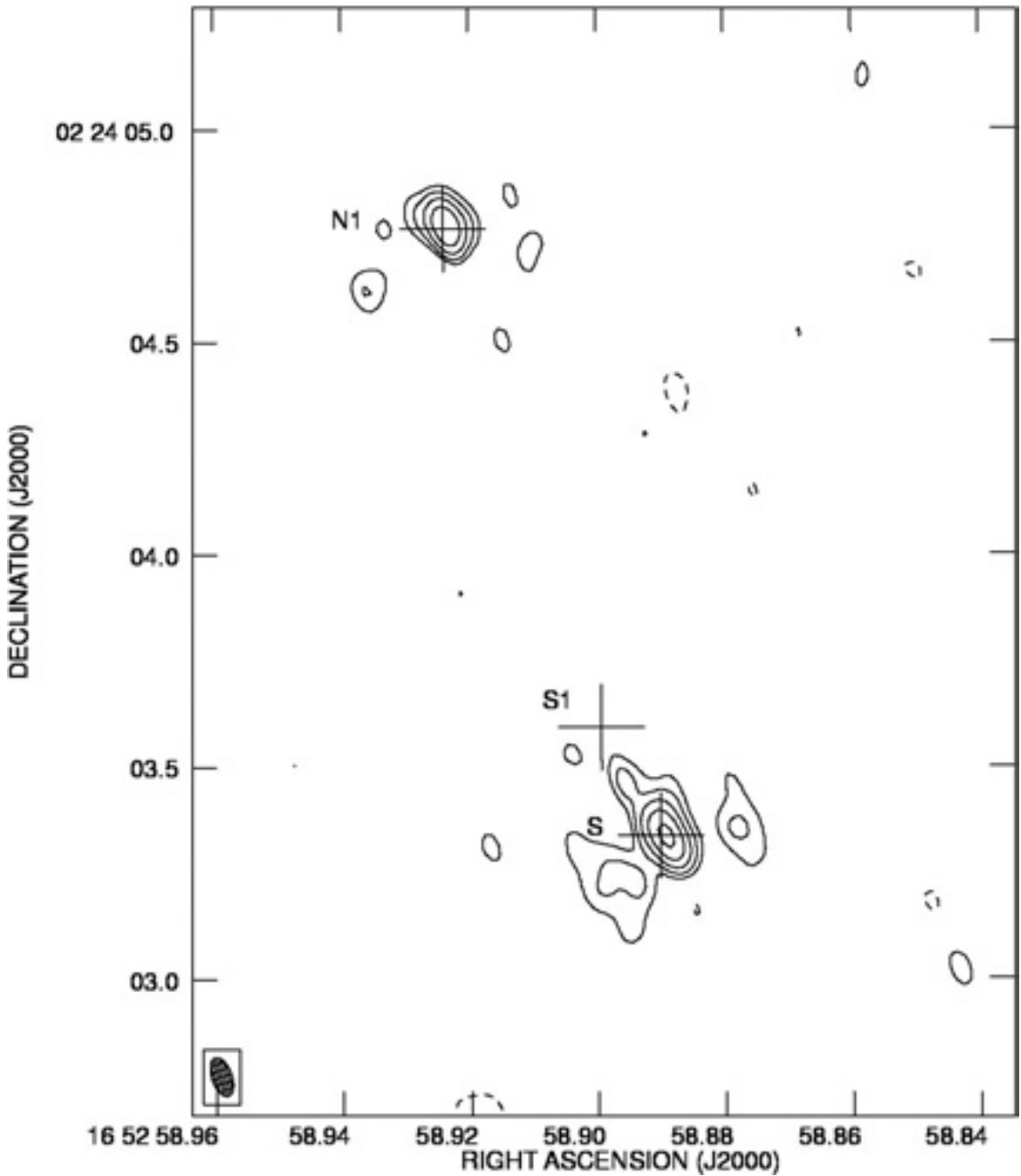

Figure 1: New MERLIN 5 GHz image of NGC 6240. The positions of the VLBA detections are marked by crosses and labels. The contour levels are 0.5 mJy beam$^{-1}$ times −1, 1, 2, 4… 32.

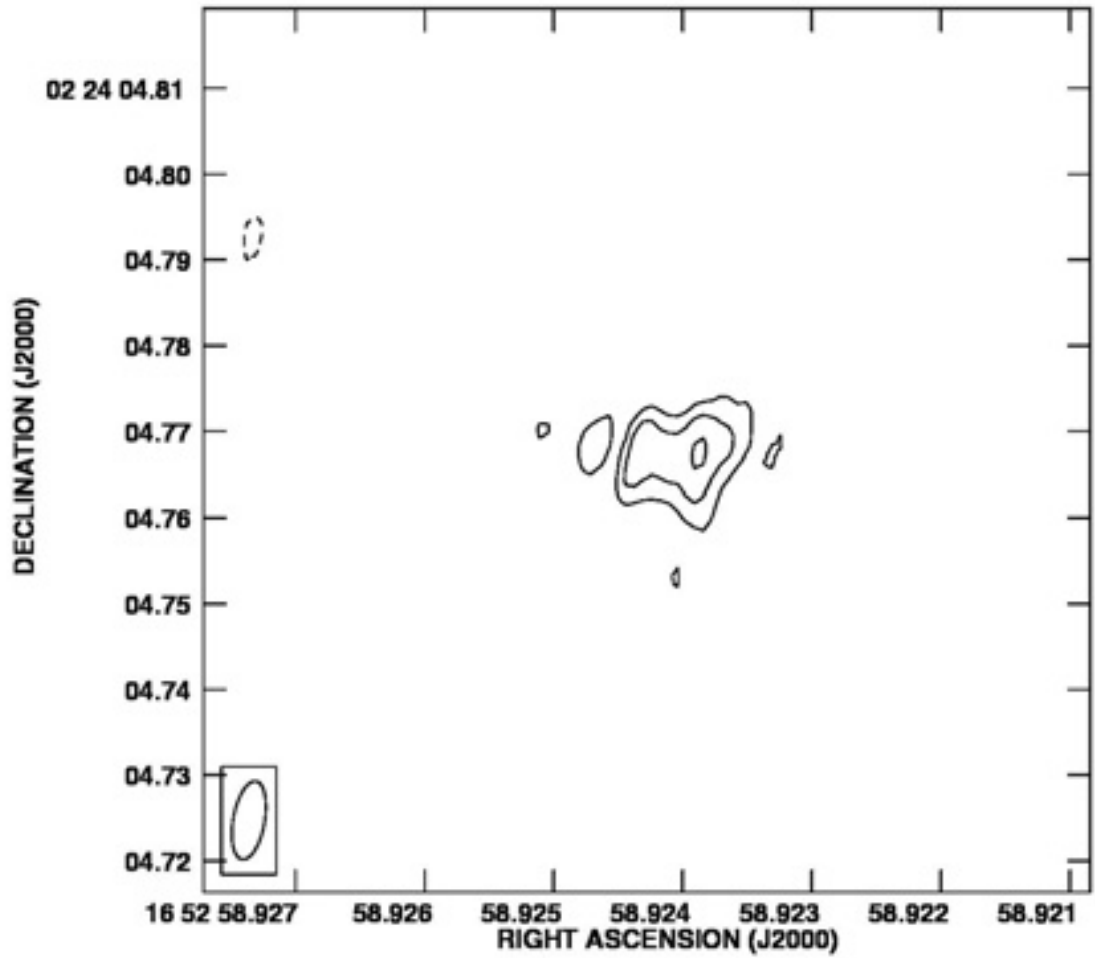

Figure 2: VLBA Images of NGC 6240 N1 at 2.4 GHz. (a) Epoch 1 image.

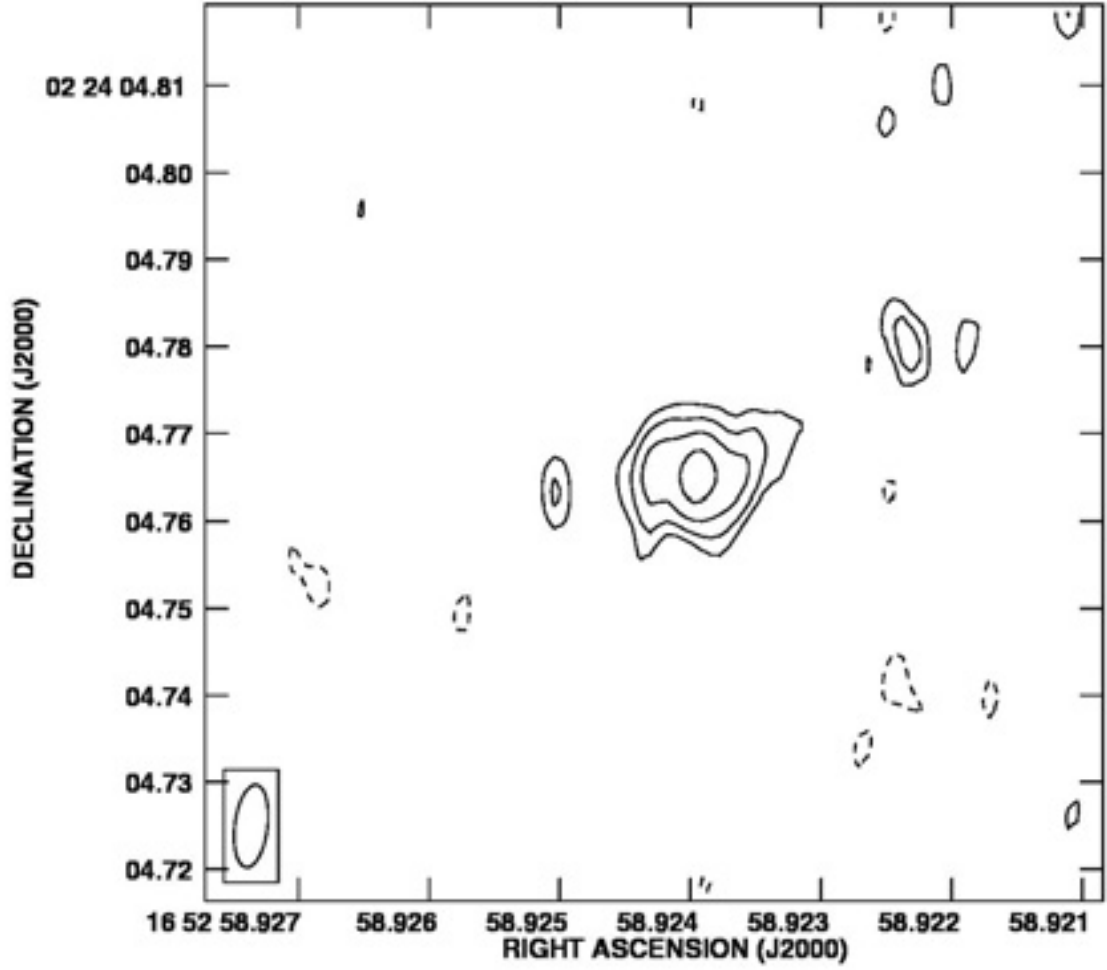

Figure 3: VLBA Images of NGC 6240 N1 at 2.4 GHz. (b) Epoch 2.

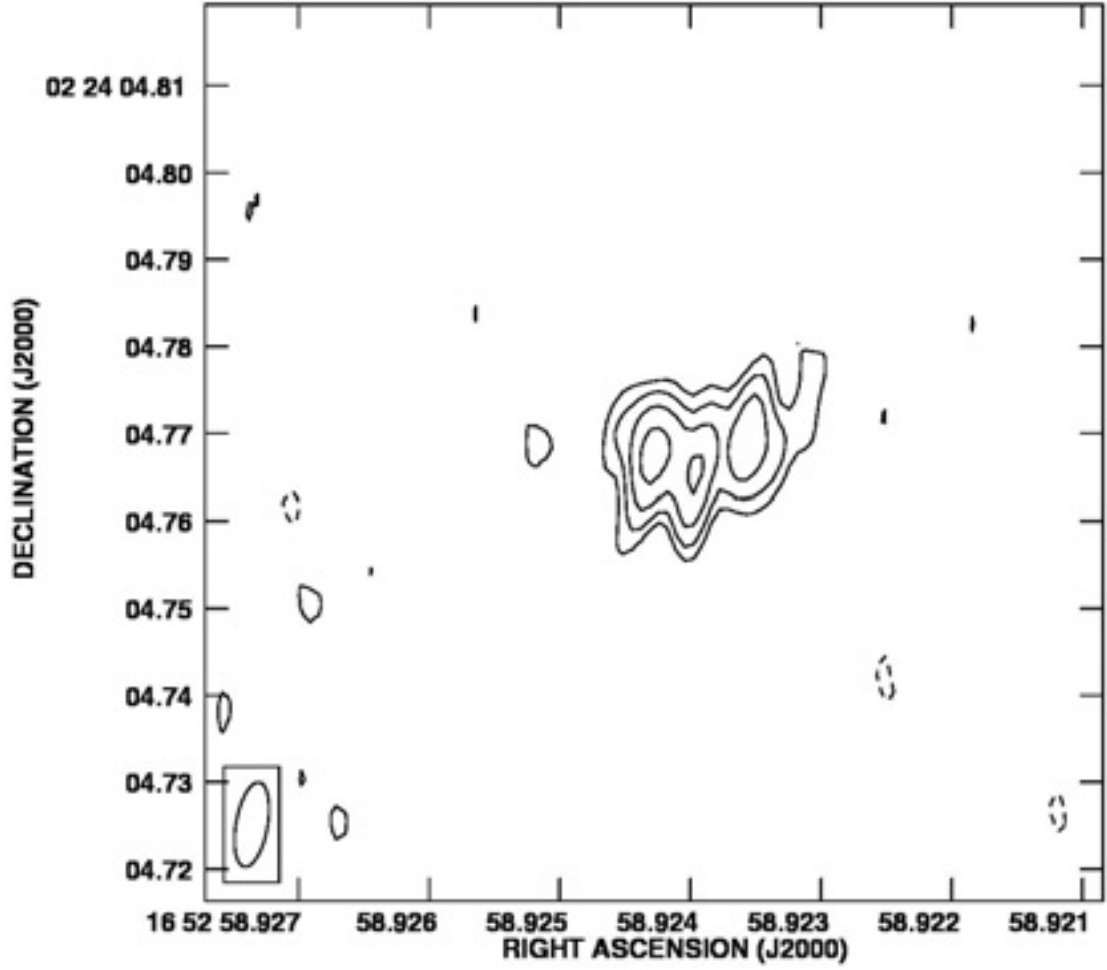

Figure 4: VLBA Images of NGC 6240 N1 at 2.4 GHz. (c) Epoch 3.

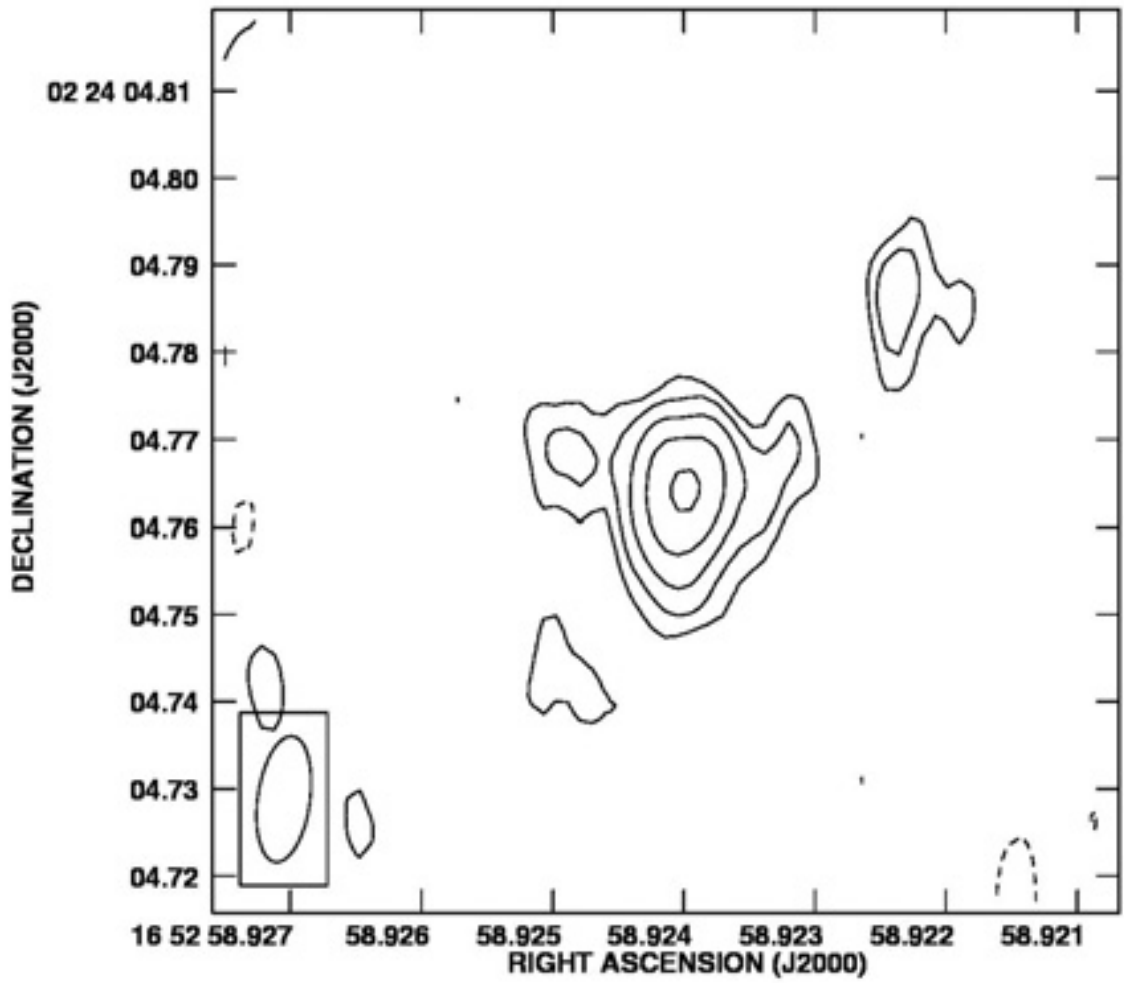

Figure 5: VLBA Images of NGC6240 N1 at 1.7 GHz. (a) Epoch 2 image.

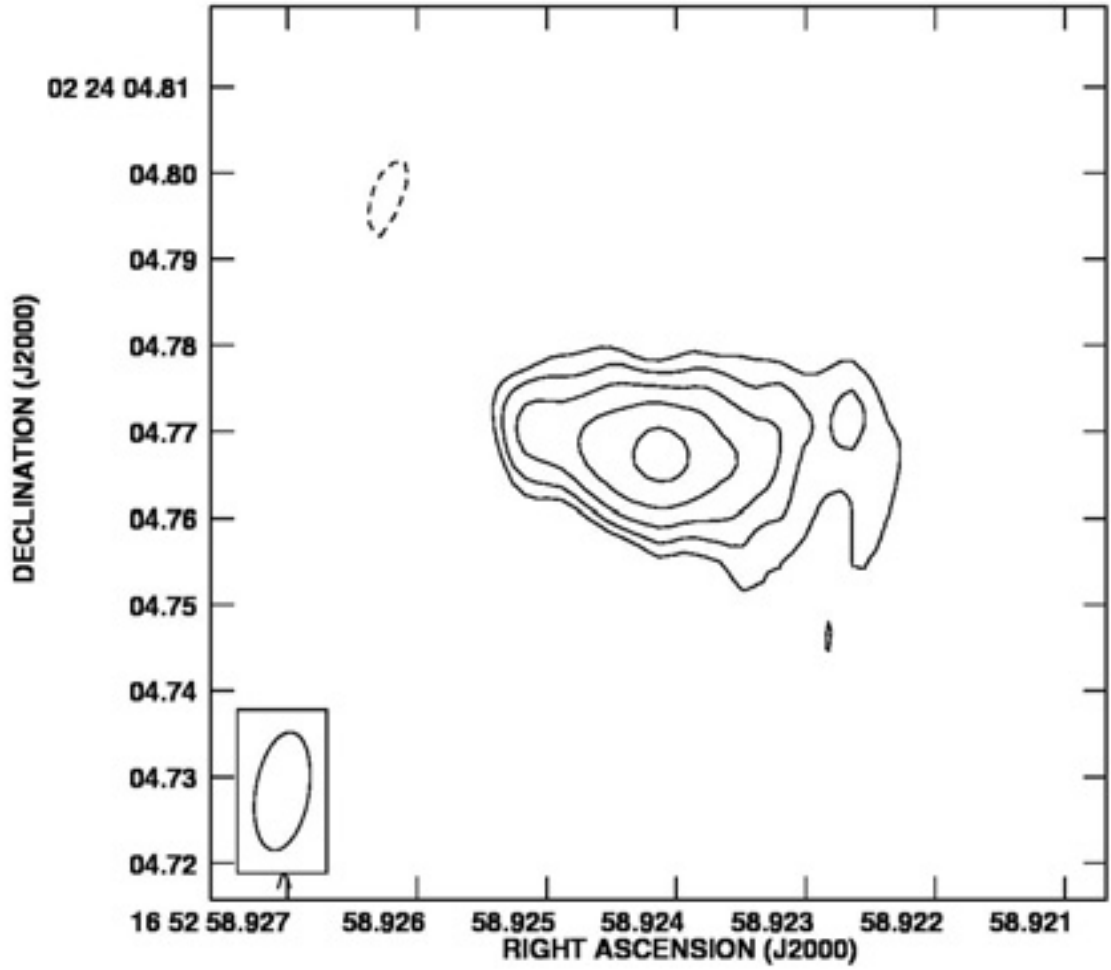

Figure 6: VLBA Images of NGC6240 N1 at 1.7 GHz. (b) Epoch 3.

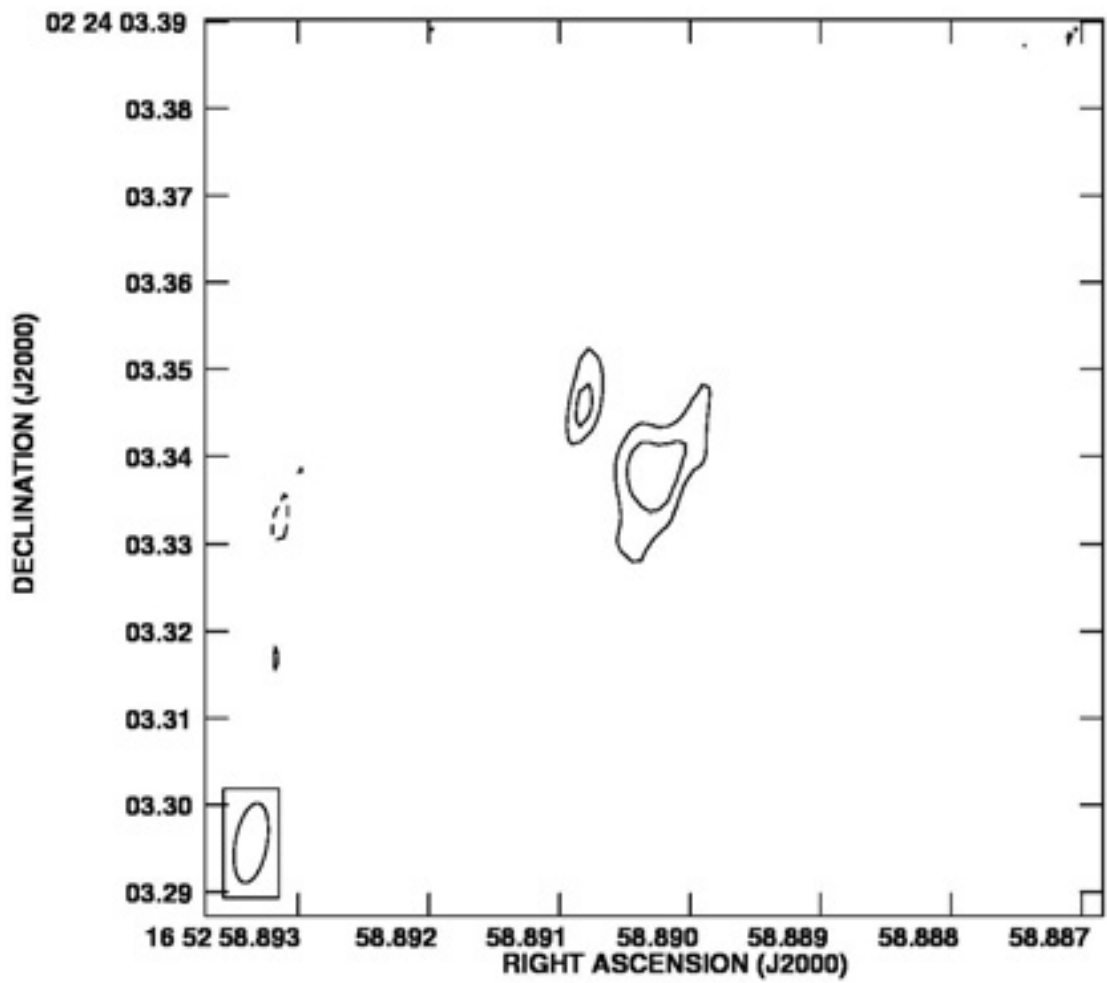

Figure 7: VLBA images of NGC6240 S at 2.4 GHz. (a) Epoch 1 image.

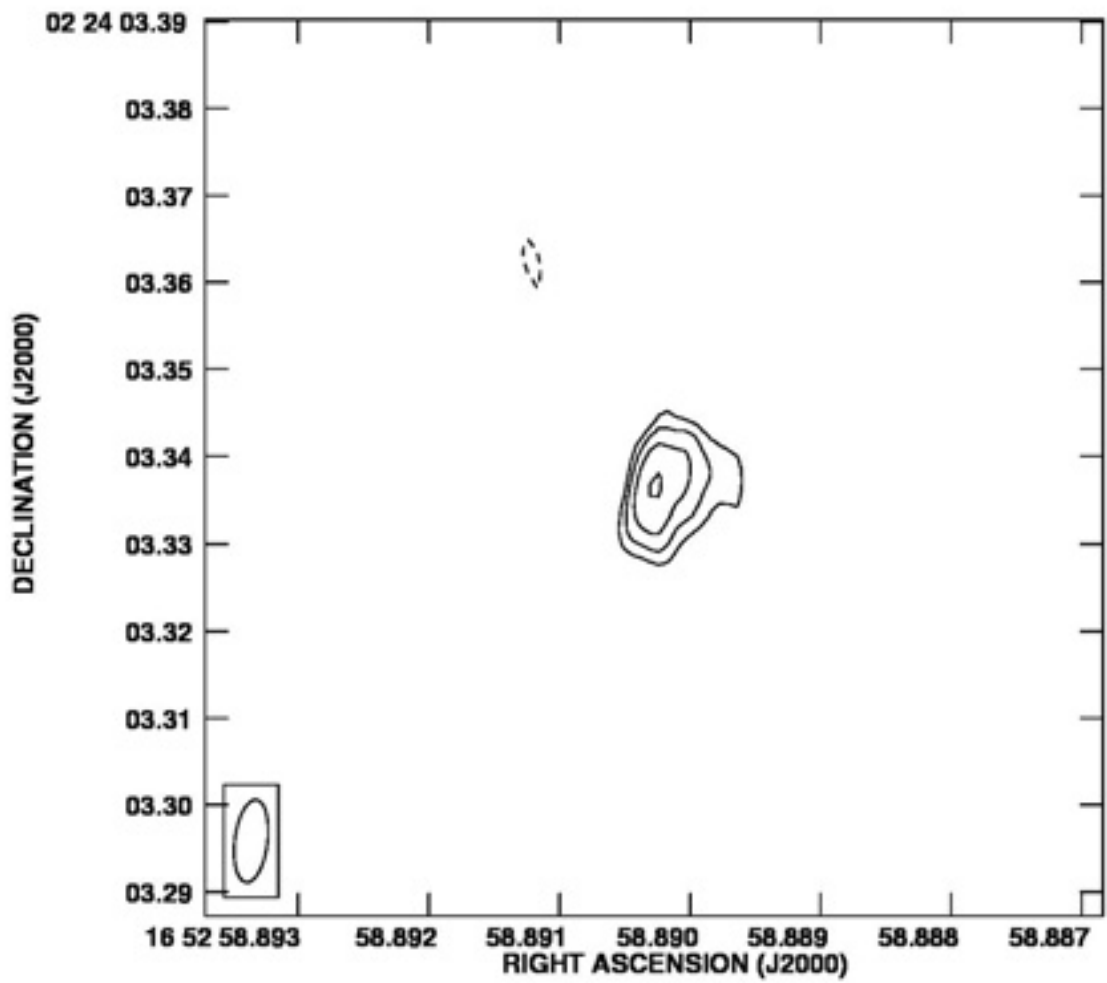

Figure 8: VLBA images of NGC6240 S at 2.4 GHz. (b) Epoch 2.

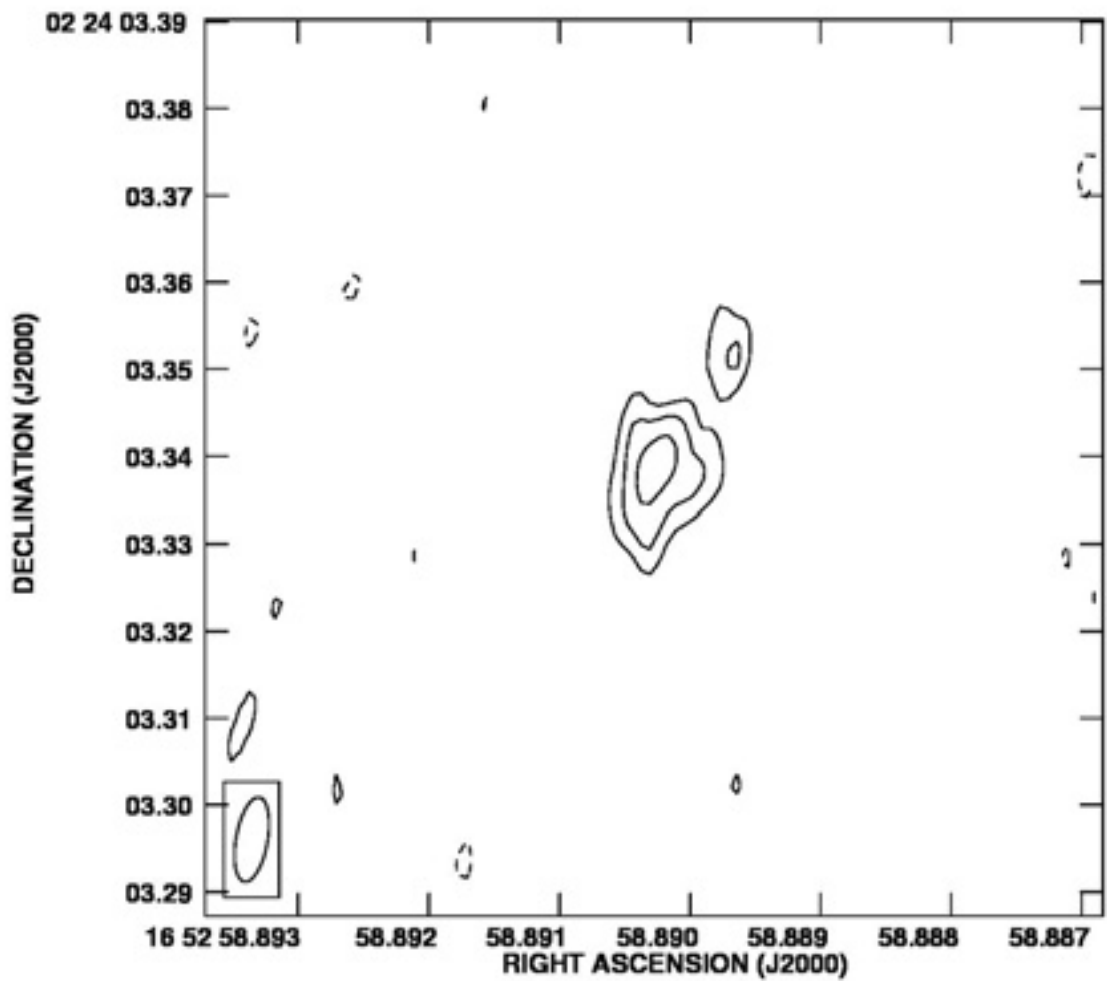

Figure 9: VLBA images of NGC6240 S at 2.4 GHz. (c) Epoch 3.

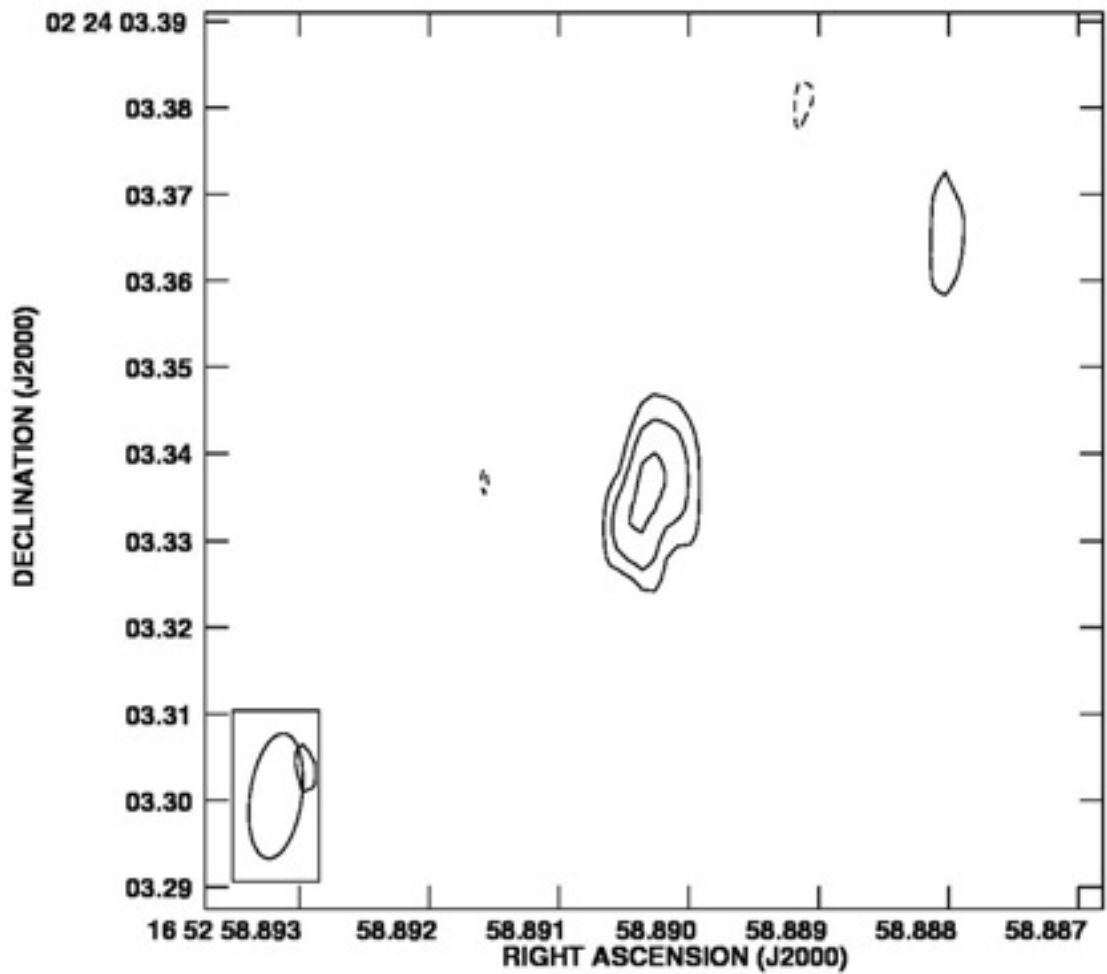

Figure 10: VLBA Images of NGC6240 S at 1.7 GHz. (a) Epoch 2 image.

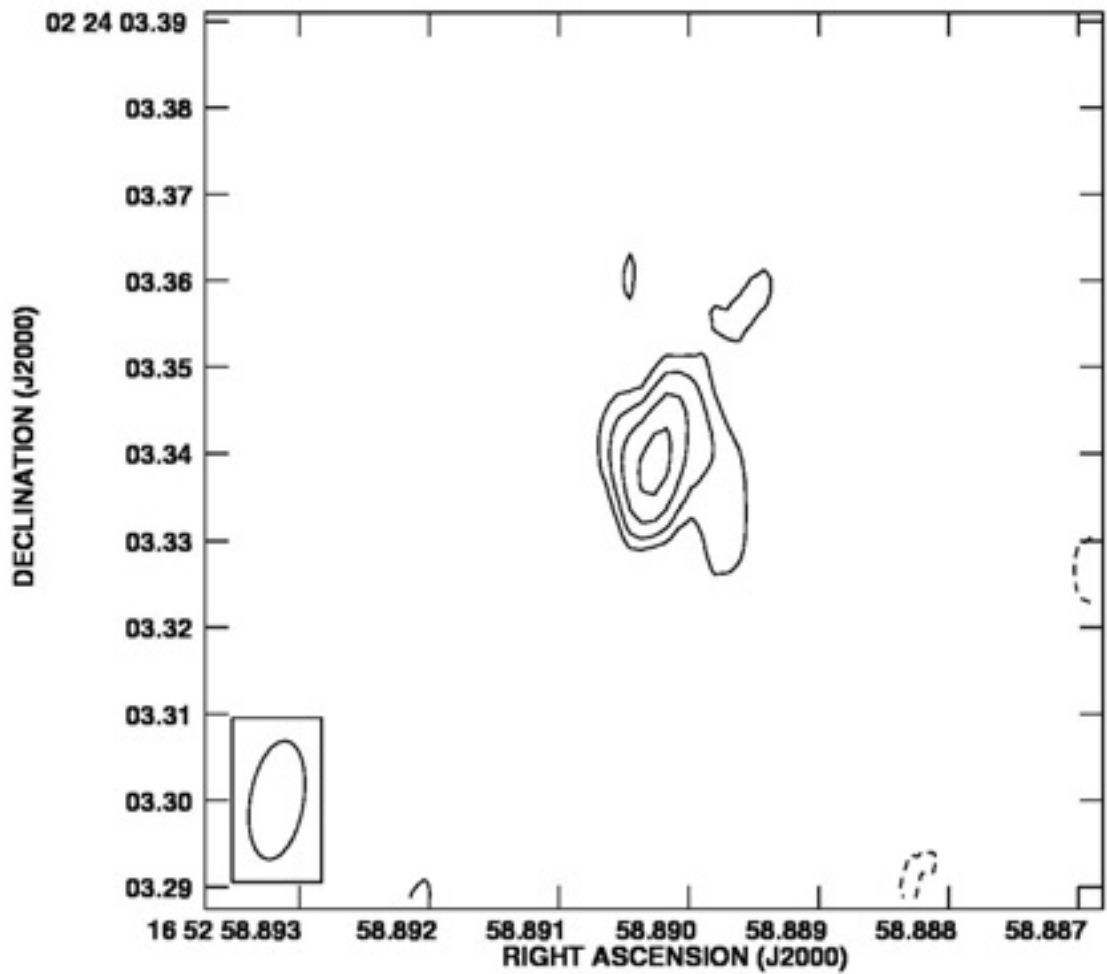

Figure 11: VLBA Images of NGC6240 S at 1.7 GHz. (b) Epoch 3.

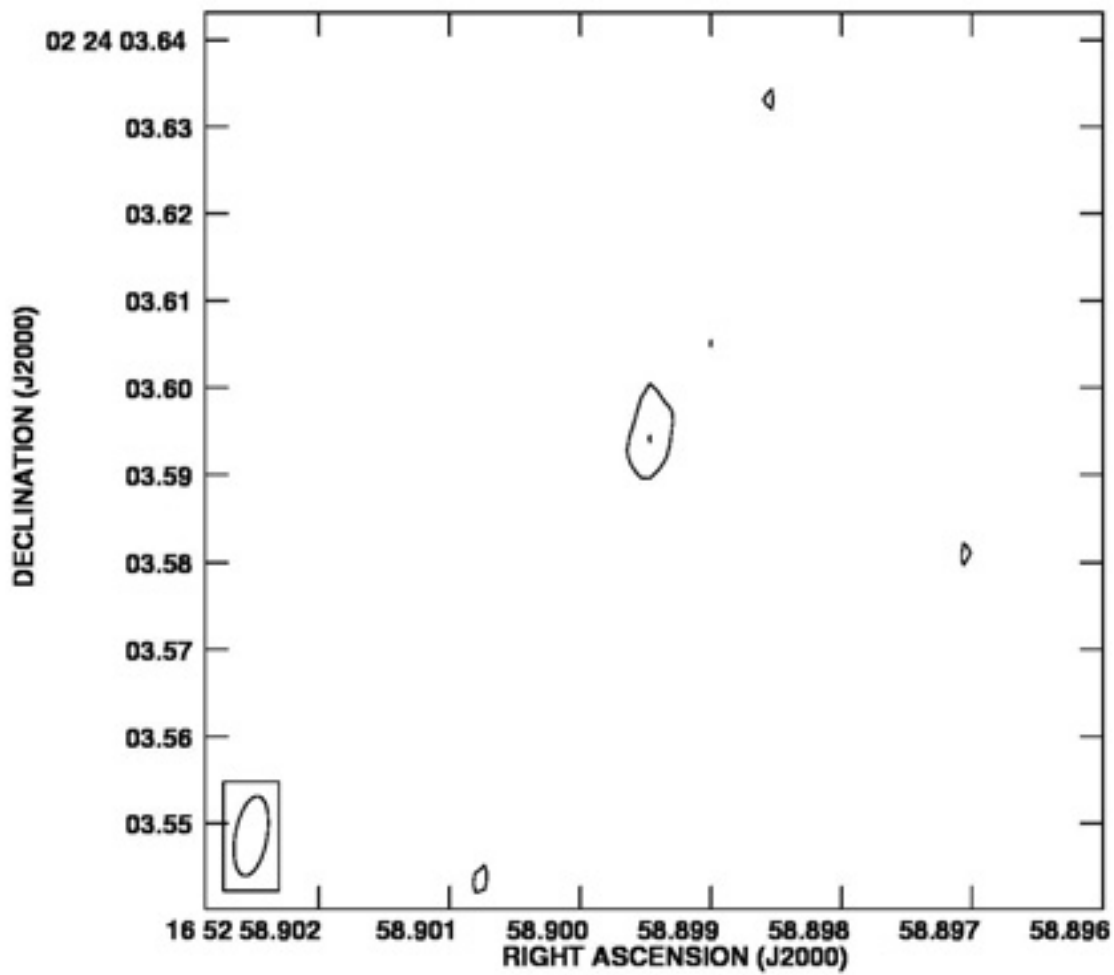

Figure 12: VLBA Images of component S1 at 2.4 GHz. (a) Epoch 1 image.

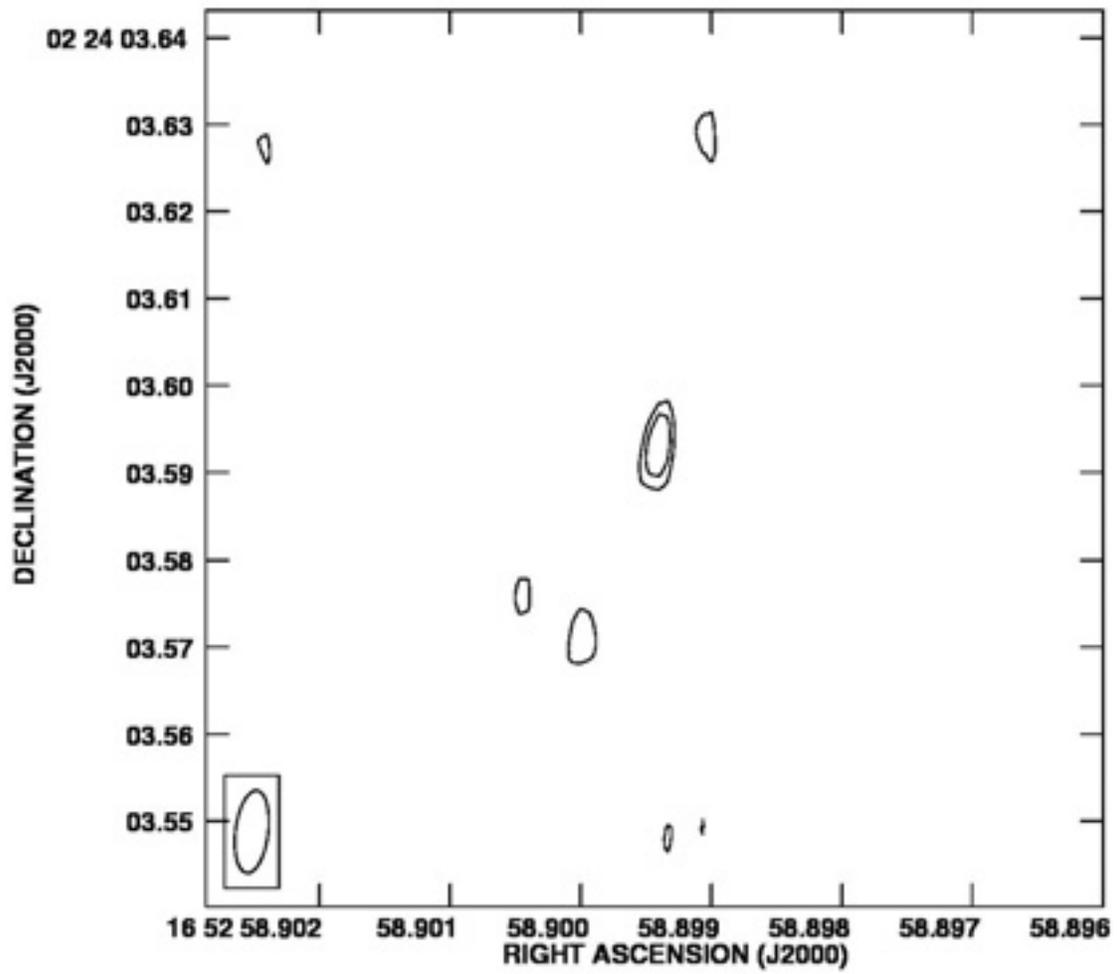

Figure 13: VLBA Images of component S1 at 2.4 GHz. (b) Epoch 2.

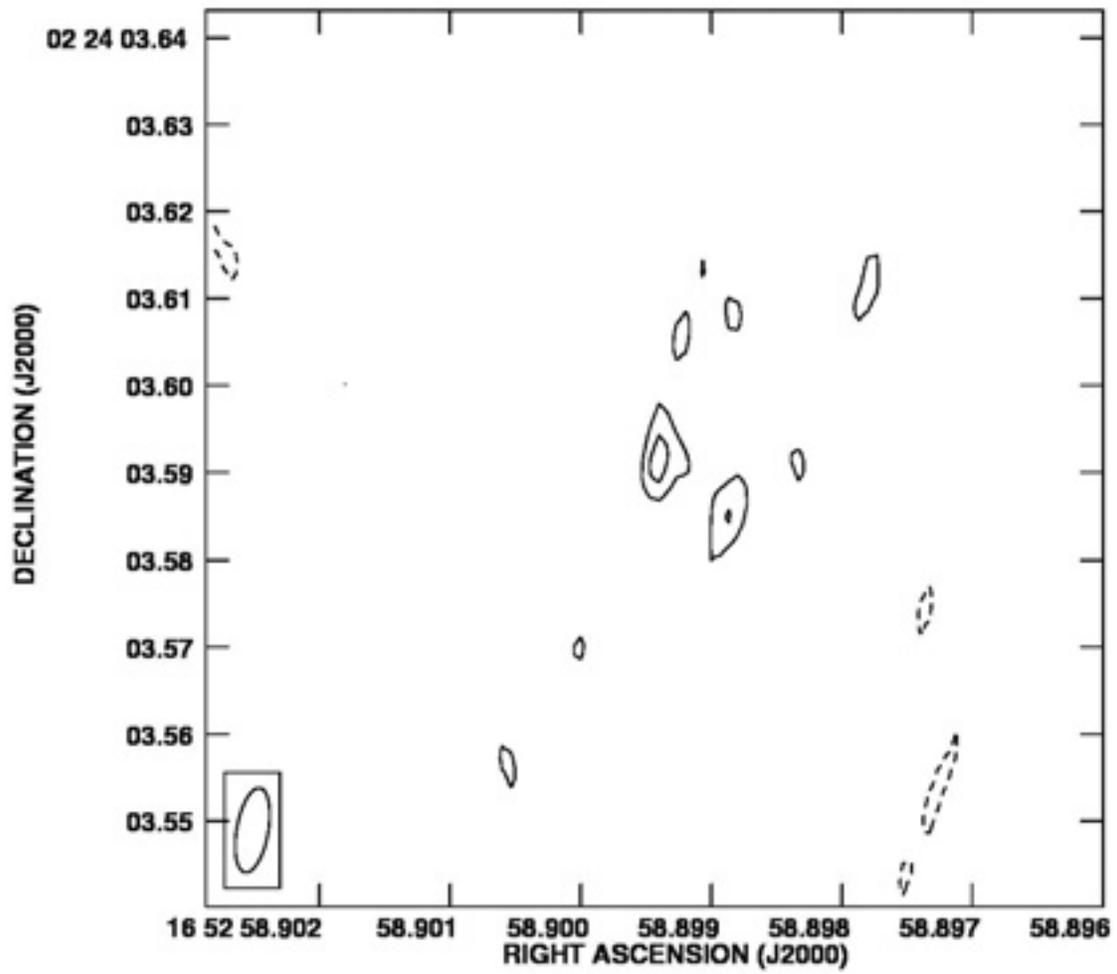

Figure 14: VLBA Images of component S1 at 2.4 GHz. (c) Epoch 3.

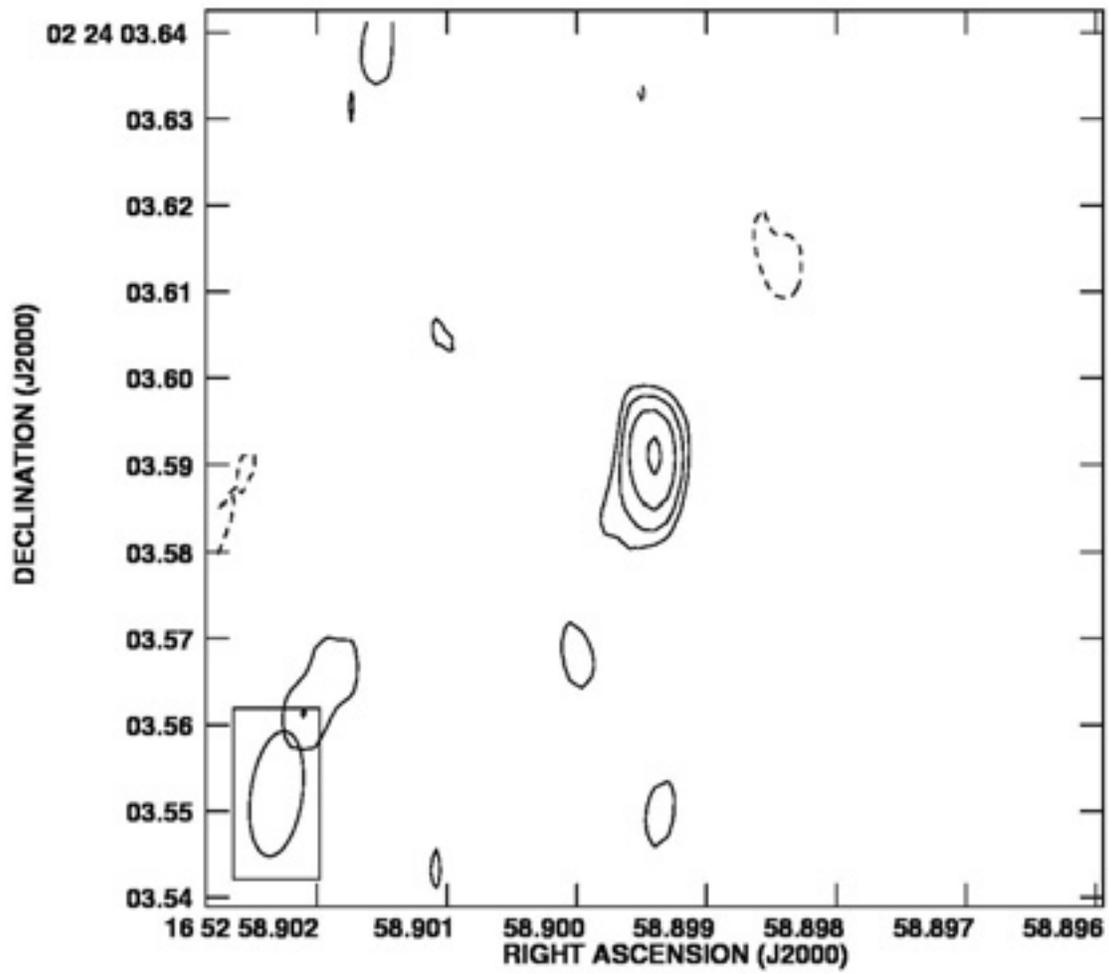

Figure 15: VLBA images of component S1 at 1.7 GHz. (a) Epoch 2 image.

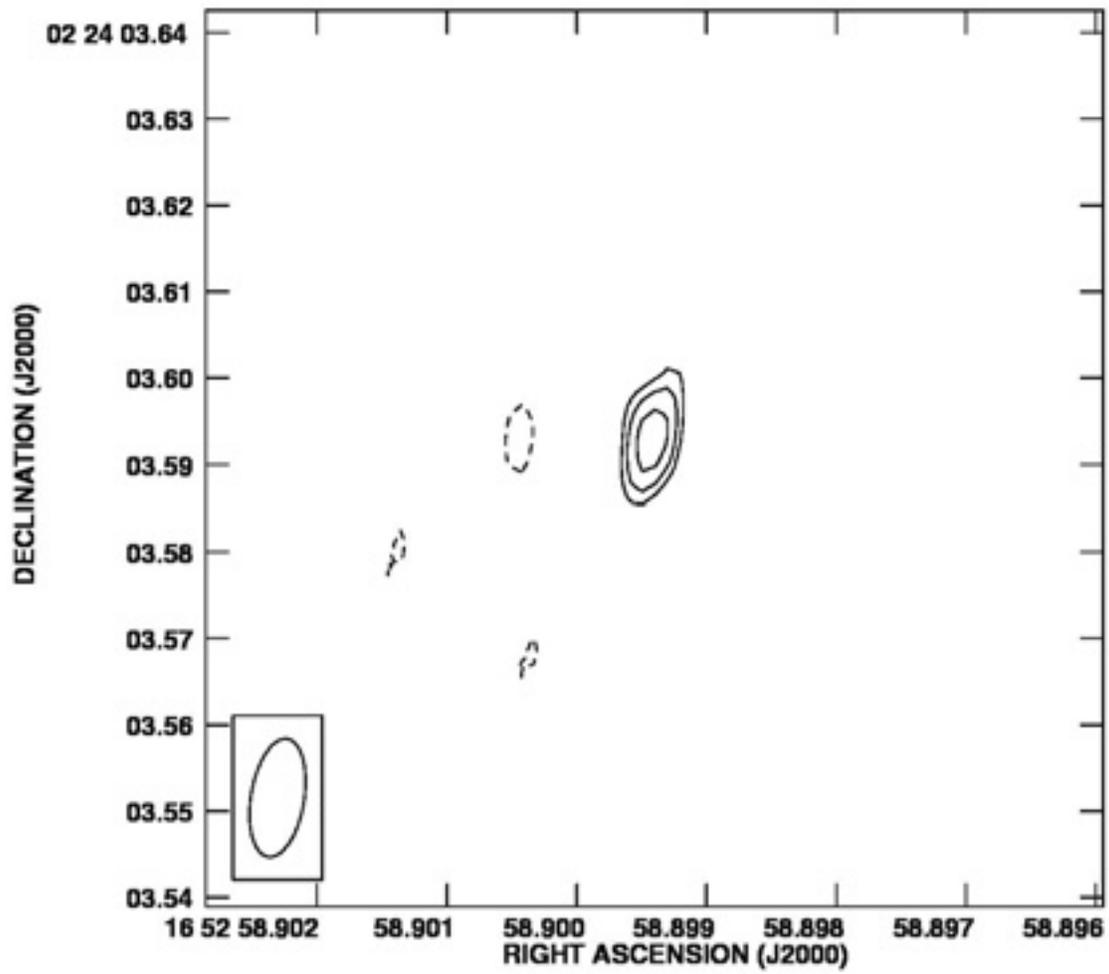

Figure 16: VLBA images of component S1 at 1.7 GHz. (b) Epoch 3.

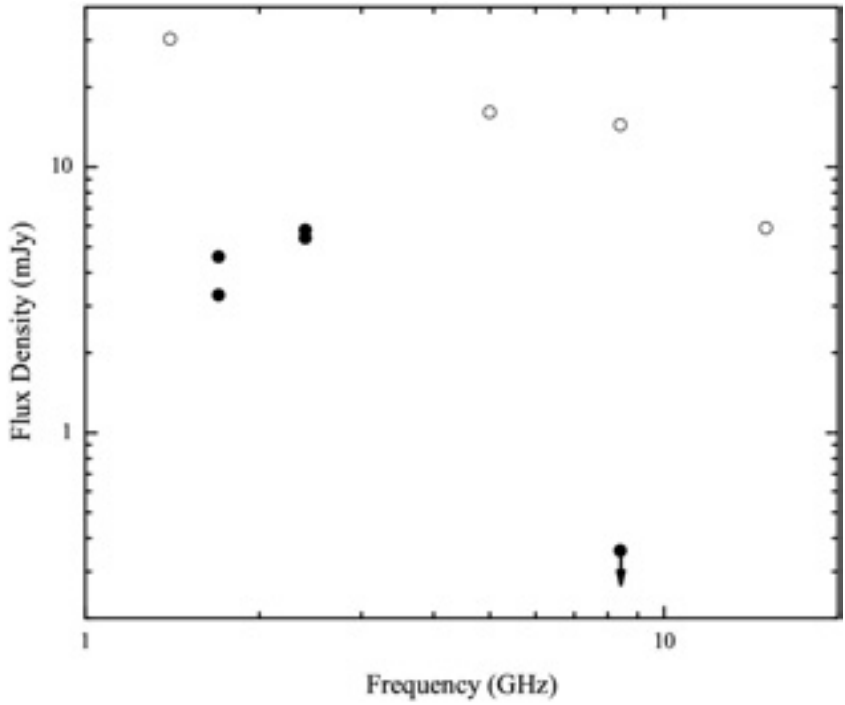

Figure 17: Radio continuum measurements of NGC6240 N1. The open circles mark MERLIN and VLA measurements (Carral et al. 1990; Colbert, Wilson, & Bland-Hawthorn 1993; Beswick et al. 2001), and the filled circles mark the VLBA measurements presented in this work.

p. 33

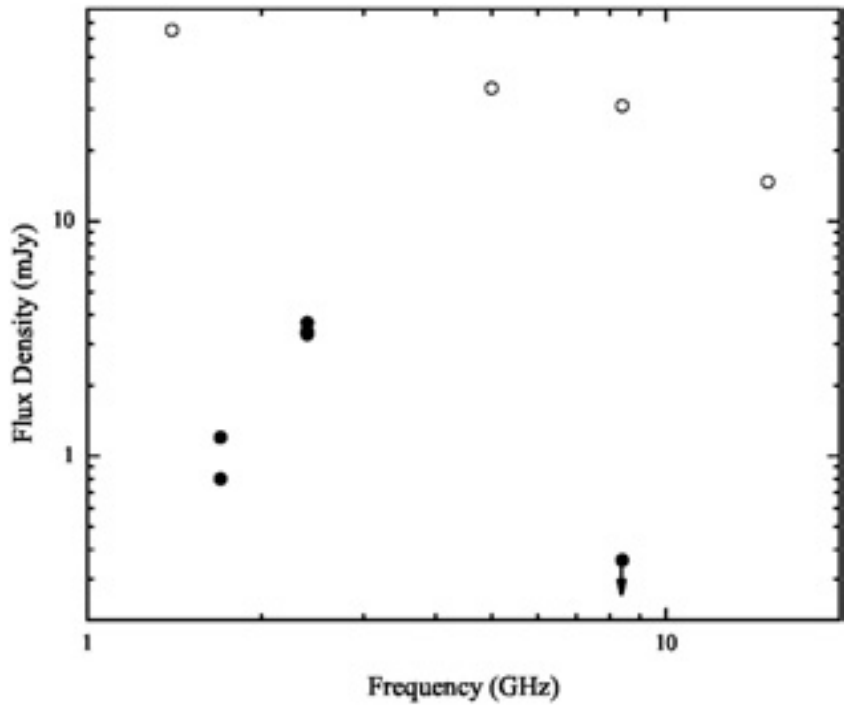

Figure 18: Radio continuum measurements of NGC6240 S. The open circles mark MERLIN and VLA measurements (Carral et al. 1990; Colbert et al. 1993; Beswick et al. 2001), and the filled circles mark the VLBA measurements presented in this work.

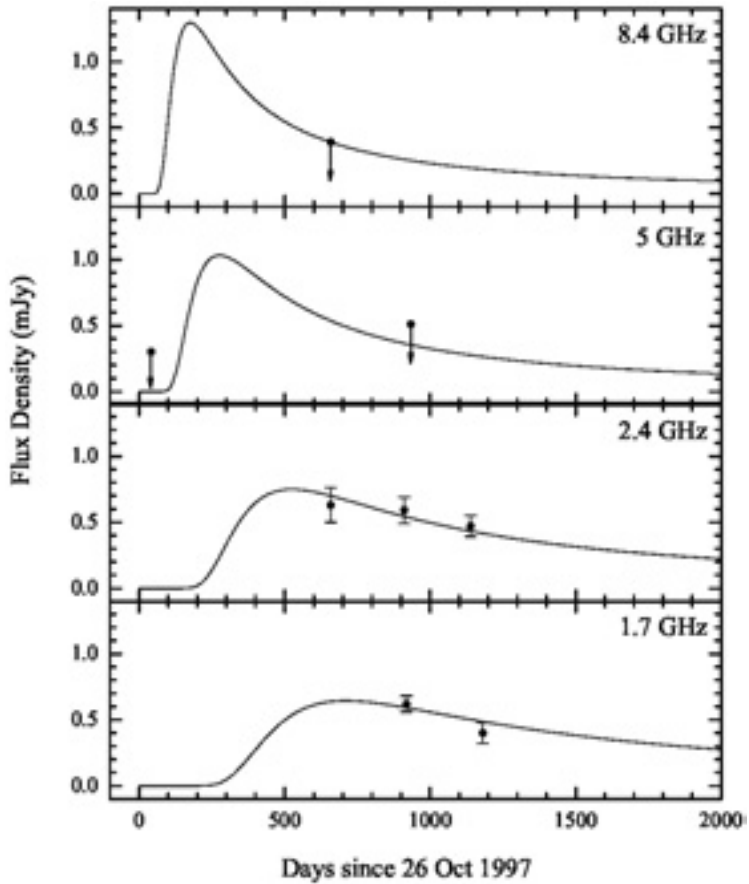

Figure 19: Radio light curves of component S1. The data are marked by filled circles, and the best-fit radio supernova model is plotted as a continuous line. Upper limits are indicated by arrows. The 5 GHz data points represent MERLIN measurements, and all other data are from the VLBA.

## Tables

### Table 1: Summary of the VLBA Observations of NGC 6240.

| Epoch | Date | Freq (GHz) | BW (MHz) | On-source Integration (min) |
|---|---|---|---|---|
| 1 | 16 Aug 1999 | 2.3 | 32 | 269 |
| 1 | 16 Aug 1999 | 8.4 | 32 | 269 |
| 2 | 4 Apr 2000 | 2.3 | 64 | 191 |
| 2 | 2 May 2000 | 1.7 | 64 | 189 |
| 3 | 8 Dec 2000 | 2.3 | 64 | 191 |
| 3 | 18 Jan 2001 | 1.7 | 64 | 189 |

### Table 2 VLBA Image Properties

| Epoch | Band (GHz) | RMS (mJy beam$^{-1}$) | Beam Size (mas × mas, deg) | | |
|---|---|---|---|---|---|
| 1 | 8.4 | 0.13 | 2.79 | 1.06 | −14.8 |
| 1 | 2.4 | 0.19 | 9.28 | 3.66 | −10.0 |
| 2 | 2.4 | 0.10 | 9.55 | 3.76 | −7.0 |
| 2 | 1.7 | 0.06 | 14.56 | 5.98 | −7.9 |
| 3 | 2.4 | 0.09 | 9.85 | 3.63 | −10.0 |
| 3 | 1.7 | 0.06 | 13.82 | 6.16 | −9.3 |

### Table 3 Contour Levels for the VLBA Images of NGC 6240

| Component | Band | Epoch | Figure | Contour Levels (mJy beam$^{-1}$) |
|---|---|---|---|---|
| N1 | S | 1 | 2 | ±0.41, 0.62, 0.92 |
| N1 | S | 2 | 2 | ±0.30, 0.44, 0.66, 1.00 |
| N1 | S | 3 | 2 | ±0.28, 0.42, 0.63, 0.95 |
| N1 | L | 2 | 3 | ±0.19, 0.29, 0.43, 0.64, 0.96 |
| N1 | L | 3 | 3 | ±0.18, 0.27, 0.41, 0.61, 0.91 |
| S | S | 1 | 4 | ±0.40, 0.60 |
| S | S | 2 | 4 | ±0.31, 0.47, 0.70, 1.05 |
| S | S | 3 | 4 | ±0.31, 0.46, 0.69, 1.03 |
| S | L | 2 | 5 | ±0.18, 0.27, 0.41 |
| S | L | 3 | 5 | ±0.17, 0.26, 0.38, 0.57 |
| S1 | S | 1 | 6 | ±0.41, 0.62 |
| S1 | S | 2 | 6 | ±0.31, 0.47 |
| S1 | S | 3 | 6 | ±0.31, 0.47 |
| S1 | L | 2 | 7 | ±0.17, 0.26, 0.38, 0.57 |
| S1 | L | 3 | 7 | ±0.18, 0.27, 0.41 |

**Table 4 Image Analysis of the Northern Nucleus (N1)**

| Epoch | Band  | $S_\nu$ | Err   | Major Axis | Minor Axis | PA  | RA       | Dec    | Err   |
|-------|-------|---------|-------|------------|------------|-----|----------|--------|-------|
|       | (GHz) | (mJy)   | (mJy) | (mas)      | (mas)      | (°) | (s)      | (')    | (mas) |
| 1     | 2.4   | 5.8     | 0.2   | 23         | 10         | 85  | 58.92405 | 4.7671 | 0.3   |
| 2     | 2.4   | 5.4     | 0.1   | 18         | 8          | 93  | 58.92403 | 4.7651 | 0.6   |
| 3     | 2.4   | 5.4     | 0.1   | 17         | 9          | 100 | 58.92397 | 4.7675 | 0.5   |
| 2     | 1.7   | 3.8     | 0.1   | 21         | 10         | 88  | 58.92399 | 4.7644 | 0.6   |
| 3     | 1.7   | 4.2     | 0.1   | 30         | 8          | 82  | 58.92403 | 4.7677 | 1.2   |

**Table 5 Image Analysis of the Southern Nucleus (S)**

| Epoch | Band  | $S_\nu$ | Err   | RA       | Dec    | Err   |
|-------|-------|---------|-------|----------|--------|-------|
|       | (GHz) | (mJy)   | (mJy) | (s)      | (")    | (mas) |
| 1     | 2.4   | 3.7     | 0.3   | 58.89030 | 3.3392 | 0.3   |
| 2     | 2.4   | 3.3     | 0.2   | 58.89014 | 3.3364 | 0.7   |
| 3     | 2.4   | 3.4     | 0.1   | 58.89021 | 3.3381 | 0.3   |
| 2     | 1.7   | 0.8     | 0.1   | 58.89027 | 3.3362 | 0.4   |
| 3     | 1.7   | 1.2     | 0.1   | 58.89021 | 3.3389 | 1.6   |

**Table 6 Image Analysis of Component S1**

| Epoch | Band  | $S_\nu$ | Err   | RA       | Dec    | Err   |
|-------|-------|---------|-------|----------|--------|-------|
|       | (GHz) | (mJy)   | (mJy) | (s)      | (")    | (mas) |
| 1     | 2.4   | 0.63    | 0.13  | 58.89949 | 3.5955 | 0.4   |
| 2     | 2.4   | 0.59    | 0.10  | 58.89940 | 3.5930 | 0.3   |
| 3     | 2.4   | 0.47    | 0.08  | 58.89940 | 3.5913 | 0.3   |
| 2     | 1.7   | 0.62    | 0.06  | 58.89944 | 3.5901 | 0.4   |
| 3     | 1.7   | 0.40    | 0.08  | 58.89943 | 3.5928 | 0.4   |

**Table 7 Supernova Light Curve Model for Component S1**

| Parameter | Value ± Error |
|---|---|
| $t_0$ | 26 Oct 1997 ± 170 days |
| $S_0$ | 2.6 ± 1.0 Jy |
| $\tau_0$ | $(3.9 \pm 0.5) \times 10^5$ |
| $\alpha$ | −0.7 (fixed) |
| $\beta$ | −1.3 (fixed) |